\documentclass{article}

\usepackage{PRIMEarxiv}

\usepackage[utf8]{inputenc} 
\usepackage[T1]{fontenc}    
\usepackage{hyperref}       
\usepackage{url}            
\usepackage{booktabs}       
\usepackage{amsfonts}       
\usepackage{nicefrac}       
\usepackage{microtype}      
\usepackage{lipsum}
\usepackage{fancyhdr}       
\usepackage{graphicx}       
\graphicspath{{media/}}     
\usepackage{amsmath}

\title{COVID-19 Detection from Pulmonary CT Scans Using a Novel EfficientNet with Attention Mechanism

}

\author{
  Ramy Farag$^{1, }$\thanks{These authors contributed equally to this work.}  , 
  Parth Upadhyay$^{2,*}$, 
  Yixiang Gao, 
  Jacket Demby's, \\
  \textbf{Katherin Garces Montoya, 
  Seyed Mohamad Ali Tousi, 
  Gbenga Omotara, 
  G. N. DeSouza$^{3}$} \\
  Vision-Guided and Intelligent Robotics (ViGIR) Laboratory \\
  Department of Electrical Engineering and Computer Science (EECS)\\
  University of Missouri- Columbia \\
  Columbia, MO, 65211\\
  \texttt{\{rmf3mc$^1$, desouzag$^3$\}@missouri.edu; pcuv48@mail.missouri.edu$^2$}
}

\begin{document}
\maketitle

\begin{abstract}
Manual analysis and diagnosis of COVID-19 through the examination of Computed Tomography (CT) images of the lungs can be time-consuming and result in errors, especially given high volume of patients and numerous images per patient.  So, we address the need for automation of this task by developing a new deep learning model-based pipeline. Our motivation was sparked by the CVPR Workshop on "Domain Adaptation, Explainability and Fairness in AI for Medical Image Analysis", more specifically, the "COVID-19 Diagnosis Competition (DEF-AI-MIA COV19D)" under the same Workshop. This challenge provides an opportunity to assess our proposed pipeline for COVID-19 detection from CT scan images. The same pipeline incorporates the original EfficientNet, but with an added Attention Mechanism: EfficientNet-AM. Also, unlike the traditional/past pipelines, which relied on a pre-processing step, our pipeline takes the raw selected input images without any such step, except for an image-selection step to simply reduce the number of CT images required for training and/or testing. Moreover, our pipeline is computationally efficient, as, for example, it does not incorporate a decoder for segmenting the lungs. It also does not combine different backbones nor combine RNN with a backbone, as other pipelines in the past did. Nevertheless, our pipeline still outperforms all approaches presented by other teams in last year's instance of the same challenge, at least based on the validation subset of the competition dataset.

\end{abstract}


\section{Introduction}

The recent COVID pandemic has affected people differently in terms of severity. In order to move from suspected to confirmed COVID-19 cases, initial diagnosis and management often relied on pulmonary images from a Computed Tomography (CT) scan, as it remains the most effective diagnostic tool. However, the inefficacy of manual analysis becomes obvious when thousands of individual CT scans need to be processed in short periods of time, which happens especially during a pandemic, with physicians having to assess numerous patients daily. In that case, the need to automate the diagnostic process becomes as obvious as the potential for mistakes to happen. Thankfully, the emergence of deep learning models in recent years has enabled their use as supportive tools in clinical diagnosis and examination.

Numerous studies in recent years have applied deep learning to CT scan image analysis. Zhang et al.(2022) utilized the VGG19 architecture. They incorporated the GlobalMax-Pool 2D Layer, achieving good performance in various metrics like sensitivity, specificity, and accuracy relative to traditional CNN models and vision transformer (ViT) models~\cite{zhang2022deep}. Zhang et al.(2021) introduced a transformer-based framework for automated COVID-19 diagnosis, comprising two primary stages: initial lung segmentation using UNet, followed by classification~\cite{zhang2021transformer}, which performed better with respect to other state-of-the-art methods. Li et al.(2020) introduced a 2D CNN for extracting features from individual slices in a CT scan, with subsequent fusion of slice-level features through a max-pooling layer and achieved improved performance in classifying COVID-19 from CT scan images and achieved 0.95 area under the receiver operating characteristic curve (AUC)~\cite{li2020using}. In the study by Shi et al.(2021), an attention mechanism encompassing both channel-wise attention (CA) and depth-wise attention (DA) was incorporated into a modified 3D Resnet18, achieving 0.99 AUC. These studies underscore the effectiveness of deep learning models in COVID-19 CT image classification tasks~\cite{shi2021dual}.

Kolliaz et al. introduced the COVID-19-CT-DB dataset~\cite{kollias2023ai, arsenos2023data, kollias2023deep, kollias2022ai, arsenos2022large, kollias2021mia, kollias2020deep, kollias2020transparent, kollias2024domain}, providing a substantial collection of labelled COVID-19 and non-COVID-19 data to tackle the demand for extensive training data in deep learning models. However, the challenge to the architecture of deep learning models arises from the diverse resolutions and numbers of slices in CT images, which depend on the imaging machine.

A few approaches have been suggested to tackle this issue. Chen et al.(2021) proposed two methods to assess CT scan image slice importance~\cite{chen2021adaptive}. First, the 2D method, Adaptive Distribution Learning with Statistical Hypothesis Testing (ADLeaST), integrates statistical analysis with deep learning for COVID-19 CT scan image classification, ensuring stable predictions by mapping images to a specific distribution. However, 2D detection is influenced by positive slices without clear symptoms during training. The 3D method incorporates self-attention structures (Within-Slice-Transformer, Between-Slice-Transformer) into the 3D CNN architecture. Still, it faces the challenges of insufficient training samples and overfitting due to its large model architecture. Hsu et al.(2023) improved this method by incorporating multi-model ensemble method~\cite{hsu2023strong}, and we adopted their approach in terms of the image selection step and the use of the EfficientNet in our proposed method.

Our work consists of a pipeline involving a simple step for CT-slice selection, followed by feature extraction, and classification using a novel EfficientNet with Attention Mechanism (EfficientNet-AM). Our approach gets away with the need for segmentation of the lungs by enhancing and highlighting the regions of interest through the attention mechanism, which allows the classification task to focus on a feature map extracted from the most representative regions of the selected CT slices.

\section{Methods}
\label{sec:headings}
This section provides an overview of the processing steps: image selection, feature extraction through EfficientNet, and the added Attention Mechanism. Subsequently, the entire proposed pipeline for COVID-19 detection from pulmonary CT scan is discussed. We used the provided challenge1 and challenge2 datasets to train, but we kept the challenge1 validation set for testing. Also, we followed \cite{hsu2023strong} for partitioning the same datasets.

\subsection{Image Selection Step}
\begin{figure}[htp!]
\centering
\includegraphics[width=0.48\textwidth]{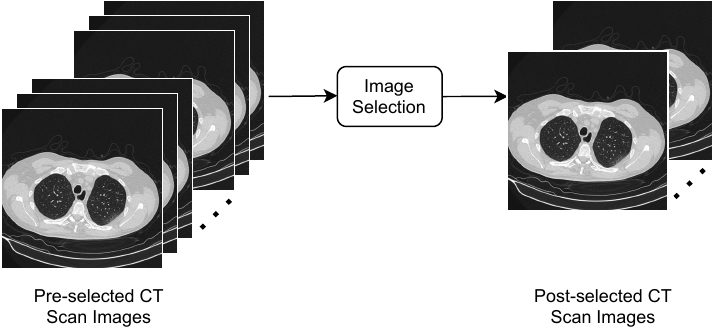}
\caption{Image selection of pulmonary CT-scan images after detection and ranking by maximum lung area. All images from a given patient are analyzed and their number is reduced to twelve or forty images for training and testing, respectively.}
\label{fig:Pre_ProcessPipeline}
\end{figure}

The goal of this step is simply to select images with maximum lung area and to reduce the number of images required for training and testing.  We followed similar steps as those presented in Hsu et al. (2023)~\cite{hsu2023strong}, and as depicted in Figure~\ref{fig:Pre_ProcessPipeline}. However, in our case, during this step, twelve CT slices are kept for training, and forty are kept for testing/validation based on the area occupied by the lungs in the images. Initially, we kept the top 50\% of CT scan slices with the largest lung area. Subsequently, we preserved twelve or forty CT scan slices, spaced equally along the already selected top 50\%, for training and testing, respectively. The rationale for keeping equally spaced CT slices is to reduce unnecessary redundancy that is expected in adjacent CT slices.

\subsection{EfficientNet}
EfficientNet represents a breakthrough in neural network architecture design by emphasizing both performance and efficiency~\cite{brock2021high}. Developed by Google AI researchers, it introduces a scalable model that achieves state-of-the-art accuracy on various computer vision tasks while maintaining a smaller number of parameters and computational cost compared to other models like ResNet or Inception~\cite{tan2019efficientnet}. 

EfficientNet achieves this by using a compound scaling method that efficiently balances model depth, width, and resolution, resulting in models that are significantly more efficient in terms of both accuracy and computational resources. This innovative approach has made EfficientNet a popular choice for various applications, from image classification to object detection and segmentation, enabling more practical deployment on resource-constrained devices without sacrificing performance. Considering these advantages, we chose to use the ECA-NFNet-L0 pre-trained model in our approach~\cite{rw2019timm}.

\subsection{Attention Mechanism}
The attention mechanism has revolutionized the field of deep learning, particularly in natural language processing and computer vision~\cite{tan2019efficientnet, vaswani2017attention}. This mechanism enhances deep learning models by allowing them to focus on relevant parts of the input data. Inspired by human cognition, it assigns varying weights to different elements based on their importance for the task. This approach helps in improving accuracy and context understanding~\cite{dosovitskiy2020image, farag2024xmnet}. In our approach, we took inspiration from the study done by Wang et al.(2021)~\cite{wang2021mask} to enhance the original EfficientNet and create the EfficientNet-AM.

\subsection{Pipeline}
\begin{figure}[ht!]
\label{fig:Pipeline}
\centering
\includegraphics[width=0.98\textwidth]{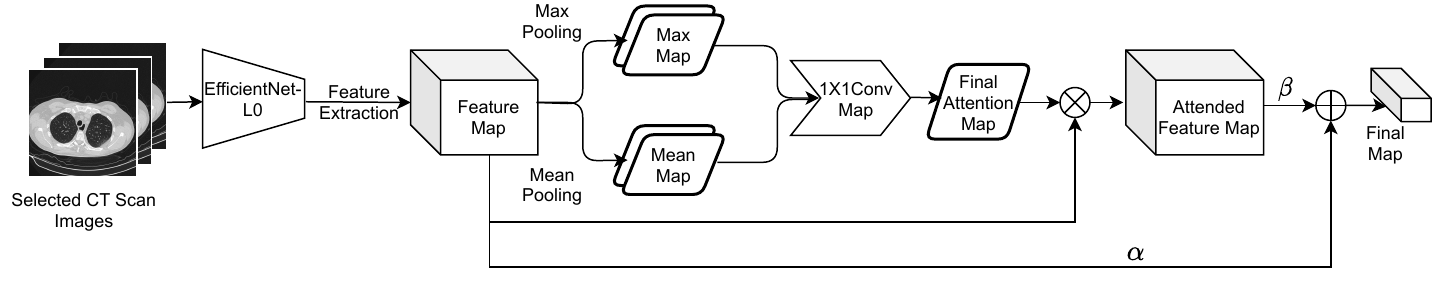}
\caption{Flowchart of our novel EfficientNet-AM method to extract features from CT scan images of lungs.}
\label{fig:Pipeline1}
\end{figure}

Figure~\ref{fig:Pipeline1} depicts the feature extraction step of our classification approach. Following the image-selection phase, an EfficientNet is utilized to extract image features, denoted as $f_{img} \subseteq R^{H \times W \times C}$, where $H$ and $W$ represent the image height and width, and $C$ is the number of channels. Next, an attention map $A_m \subseteq R^{H \times W}$ is generated by an attention module, which identifies informative parts in images. Using this attention map $A_m$ and the image features $f_{img}$, attended features $f_{att} \subseteq R^{H \times W \times C}$ are computed using spatial-wise multiplication. 

\begin{equation}
    f_{att}(i) = f_{img}(i) \circ A_m
\end{equation}
where $i$ is the channel index in $f_{img}$ and $f_{att}$, while $\circ$ represents spatial-wise multiplication.

The aim is to get the most informative parts of the image to mitigate the potential influence of false positives. Both the attended features $f_{att}$ and the original image features $f_{img}$ are leveraged to obtain merged feature $f_{merged} \subseteq R^{H \times W \times C}$.

\begin{equation}
    f_{merged}(i) = \alpha f_{img} + \beta f_{att}
\end{equation}


The equation is subject to the constraint $\alpha + \beta = 1$, where $\alpha$ and $\beta$ are hyperparameters responsible for balancing the contribution between the two feature maps. In this study, after empirical analysis, we set $\alpha = 0.5$ and $\beta = 0.5$ to equally support the importance of the original and the attended features in the Final Map.

As mentioned earlier, this combination of attended and original image features helps mitigate false positives. The final features $f_{final} \subseteq R^C$ are used for classification and are obtained by performing global average pooling across each channel. 
\begin{equation}
    f_{final} = \frac{1}{H \times W} \sum_{i=1}^H \sum_{j=1}^W f_{merged}(i,j,:)
\end{equation}

Finally, once our EfficientNet-AM generates the final feature map $f_{final}$ for the selected slice, the same $f_{final}$ features are forwarded to a dense-layer NN with a sigmoid activation function. This last-stage NN is appended at the end of the original pipeline in Figure~\ref{fig:Pipeline1}. This final-stage NN is also responsible for calculating the confidence of each selected CT-slice in being classified as \textit{positive} or \textit{negative} for COVID-19. 

\begin{figure}[htp!]
\centering
\includegraphics[width=0.80\textwidth]{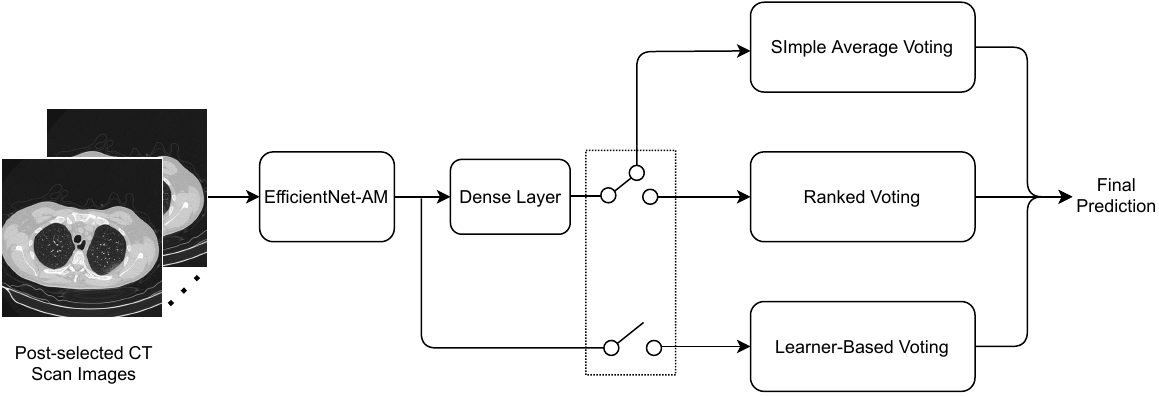}
\caption{Complete pipeline for the proposed EfficientNet-AM, for COVID-19 detection in pulmonary CT-scan images.}
\label{fig:Pipeline2}
\end{figure}

Figure~\ref{fig:Pipeline2} depicts the complete pipeline of the proposed EfficientNet-AM for COVID-19 detection from CT-scan images. Here, the longer, detailed pipeline in Figure~\ref{fig:Pipeline1} is condensed into the block "EfficientNet-AM", after which the dense NN layer for the final classification of individual samples is appended. Next, one of the three possible voting schemes (the dotted rectangle in the figure) is applied for testing/validation -- i.e. for the final diagnoses of the patients and their samples.

\subsection{Voting Schemes}
During testing, for the set of $n$ selected CT-slices from a single patient (here, $n=40$), a final diagnosis for the patient is produced based on one of three possible voting schemes: simple voting, ranked voting, or learner voting.  This "Decision Block", i.e. the block that yields the final patient-wise prediction, is shown in Figure~\ref{fig:Pipeline2}. The following subsections detail each voting scheme.

\subsubsection{Simple Average Voting}

In this scheme, the final decision involves averaging all $n$ confidences and setting a threshold $t$ at 0.5 for the COVID diagnoses. Consequently, if the average confidence exceeds this threshold, we categorize the entirety of the patient samples as COVID-19 \textit{positive}; if not, it is considered \textit{negative}.

The average confidence, $C_{\text{avg}}$, of classifying the images as COVID-19 positive is calculated as follows:
\begin{equation}
    C_{\text{avg}} = \frac{1}{n} \sum_{i=1}^{n} C_i
\end{equation}

The classification rule is defined by:
\begin{equation}
    \text{Classification} = 
    \begin{cases} 
        \text{COVID-19 Positive} & \text{if } C_{\text{avg}} > t \\
        \text{COVID-19 Negative} & \text{otherwise}
    \end{cases}
\end{equation}
where $t = 0.5$.

\subsubsection{Ranked Voting}


For ranked voting, we first sort the confidences associated with each CT-slice according to the most likely COVID-positive diagnoses. Then, ranked voting is applied to the top 2.5\% of the positive diagnoses (i.e. the most decisively towards COVID) and the bottom 2.5\% of the negative diagnoses (i.e. the most decisively towards not COVID).


\subsubsection{Learner-Based Voting}
For this final scheme, we bypass the dense NN (refer to Figure~\ref{fig:Pipeline2}) and connect the EfficientNet-AM directly to a Learner.  That is, after obtaining the feature map from the EfficientNet-AM for each selected CT-slice of the same patient, the same features are input directly into a Single-Head Attention (SHA) transformer. This SHA-Transformer, i.e. the Learner-Based Voting, will then determine whether all patient samples should be classified as COVID-19 positive or negative. 

\section{Results}
In this section, we report the results from all three voting schemes derived from the EfficientNet-AM pipeline. We also compare and evaluate how our proposed model performs against other teams' models from previous years on the validation dataset. The results of the three voting schemes are presented in Table \ref{table:example_four_column}. Here, the first column contains the type of voting, and the second and third columns provide the AUC and Macro-F1 scores obtained using the validation dataset for the respective voting scheme.
\vspace{1mm}

\begin{table}[htpb!]
\centering 
\begin{tabular}{|l|l|l|} 
\hline 
Method  & AUC & Macro-F1 \\ 
\hline 
Simple Average Voting  & 0.9713 & 93.67 \\ 
Ranked Voting  &  0.9717 &  93.35 \\ 
Learner-Based Voting  & 0.9645 &  93.67 \\
\hline 
\end{tabular}
\vspace{5mm}
\caption{Comparative results from three voting schemes derived from the EfficientNet-AM.} 
\label{table:example_four_column} 
\end{table}

Finally, we compare the performance of models proposed in previous years against the best of our three models (Simple Average Voting). These results are presented in Table \ref{table:example_table}. The Macro-F1 scores listed in Table \ref{table:example_table} are the ones reported by the respective teams' publications (see citations on the table). As the reader will notice, our model outperformed all other teams' competing models on the validation dataset.

\begin{table}[htpb!]
\centering 
\begin{tabular}{|l|l|} 
\hline 
Method & Macro-F1 \\ 
\hline 
Eff-mix-conv-E \cite{hsu2023strong} & 0.922 \\ 
EDPS-COVID-19-CT-LS \cite{turnbull2023enhanced} & 0.932\\ 
IPSR-4L-CNN-C \cite{morani2022covid} & 0.851\\
ResNet3D-18 + MHA \cite{rondinella2023unict} & 0.9021\\
\hline
\textbf{Ours}  & \textbf{0.9367}\\

\hline 
\end{tabular}
\vspace{5mm}
\caption{Comparison between our EfficientNet-AM method, using simple average voting, and previous year's approaches, using Macro-F1 score on the validation set.}
\label{table:example_table} 
\end{table}

\section{Conclusion}
In this study, we presented a pipeline with three different voting schemes after the feature extraction and classification using a novel EfficientNet with Attention Mechanism (EfficientNet-AM).  Our framework aimed at enhancing the detection of COVID-19 from pulmonary CT images. The voting mechanism included: a simple average voting, a ranked voting -- both of which are based on the confidence levels from the EfficientNet-AM classification module over a number (40) of CT scan slices -- and a Learner-based voting, which learns directly from the feature map from the same CT slices. Based on the Macro-F1 scores reported by last year's teams, we have demonstrated that our model is a likely contender for the best model for this year's challenge.
In the future, we intend to study further the balance between attended (from the attention mechanism) versus the original features (from EfficientNet) and how to provide more attention to more relevant areas of the lungs. Also, further investigation on why the Learner-based voting could not outperform the simple voting scheme is warranted. Finally, the use of EfficientNet-AM for other applications, such as leaf venation, have already started and will be expanded in order to evaluate the usefulness of this network beyond what has been demonstrated so far. 

\section*{Acknowledgements}

Computational resources for this research have been supported by the NSF National Research Platform, as part of GP-ENGINE (award OAC \#2322218).

\bibliographystyle{unsrt}  
\bibliography{references}

\end{document}